\documentclass[twocolumn]{aastex62}
\usepackage{amsmath,amstext}
\usepackage[T1]{fontenc}
\usepackage{apjfonts}
\usepackage[figure,figure*]{hypcap}

\newcommand{\runone}{A\_}
\newcommand{\runonenu}{A}

\newcommand{\runtwonu}{B}

\begin{document}

\title{Simulating Metal Mixing of Both Common and Rare Enrichment Sources in a Low Mass Dwarf Galaxy}

\correspondingauthor{Andrew Emerick}
\email{aemerick@carnegiescience.edu}

\author[0000-0003-2807-328X]{Andrew Emerick}
\altaffiliation{Carnegie Fellow in Theoretical Astrophysics}
\affiliation{Carnegie Observatories, Pasadena, CA, 91101, USA}
\affiliation{TAPIR, California Institute of Technology, Pasadena, CA, 91125, USA}
\author[0000-0003-2630-9228]{Greg L. Bryan}
\affiliation{Department of Astronomy, Columbia University, New York, NY, 10027, USA}
\affiliation{Center for Computational Astrophysics, Flatiron Institute, 162 5th Ave, New York, NY, 10010, USA}
\author[0000-0003-0064-4060]{Mordecai-Mark Mac Low}
\affiliation{Department of Astrophysics, American Museum of Natural History, New York, NY, 10024, USA}
\affiliation{Department of Astronomy, Columbia University, New York, NY, 10027, USA}
\affiliation{Center for Computational Astrophysics, Flatiron Institute, 162 5th Ave, New York, NY, 10010, USA}

\keywords{Galaxy chemical evolution---Dwarf galaxies---Chemical enrichment---Hydrodynamics}

\begin{abstract}
One-zone models constructed to match observed stellar abundance patterns have been used extensively to constrain the sites of nucleosynthesis with sophisticated libraries of stellar evolution and stellar yields. The metal mixing included in these models is usually highly simplified, although it is likely to be a significant driver of abundance evolution. In this work we use high-resolution hydrodynamics simulations to investigate how metals from individual enrichment events with varying source energies $E_{\rm ej}$ mix throughout the multi-phase interstellar medium (ISM) of a low-mass ($M_{\rm gas}=2\times 10^{6}$~M$_{\odot}$), low-metallicity, isolated dwarf galaxy. These events correspond to the characteristic energies of both common and exotic astrophysical sites of nucleosynthesis, including: asymptotic giant branch winds ($E_{\rm ej}\sim$10$^{46}$~erg), neutron star-neutron star mergers ($E_{\rm ej}\sim$10$^{49}$~erg), supernovae ($E_{\rm ej}\sim$10$^{51}$~erg), and hypernovae ($E_{\rm ej}\sim$10$^{52}$~erg). We find the mixing timescales for individual enrichment sources in our dwarf galaxy to be long (100~Myr--1~Gyr), with a clear trend of increasing homogeneity for the more energetic events. Given these timescales, we conclude that the spatial distribution and frequency of events are important drivers of abundance homogeneity on large scales; rare, low $E_{\rm ej}$ events should be characterized by particularly broad abundance distributions. The source energy $E_{\rm ej}$ also correlates with the fraction of metals ejected in galactic winds, ranging anywhere from 60\% at the lowest energy to 95\% for hypernovae. We conclude by examining how the radial position, local ISM density, and global star formation rate influence these results.
\end{abstract}

\section{Introduction}

The elemental abundances of a galaxy over time are sensitive to the nuclear physics and stellar astrophysics that determines which stars make what elements and when. The stellar abundance patterns within a given galaxy also depend on the details of how those metals are released into the interstellar medium (ISM) through various forms of stellar feedback, and the hydrodynamic interactions that ultimately mix those elements into the ISM or eject them from the galaxy in galactic winds. 

Old, metal poor systems are enriched by substantially fewer sources than younger, more metal rich stellar populations. Studying abundance patterns in these environments allows us to place constraints on the nucleosynthetic sites of each element and the total yields of individual enrichment events. Therefore, the lowest mass dwarf galaxies in the Local Group, ultra-faint dwarfs (UFDs), offer some of the best constraints on these processes. For example, these environments can potentially be used to trace and constrain enrichment from the first stars \citep[e.g.][]{FrebelBromm2012,Ji2015,Ritter2015,Jeon2017,Hartwig2018} and have been used to place constraints on the variety of possible astrophysical sources of $r$-process enrichment \citep[e.g.][]{Ji2016a,Ji2016b,Ji2018a,Ji2019,Tsujimoto2017,Duggan2018,Nagasawa2018,Ojima2018,Skuladottir2019}.


The stochastic sampling of individual enrichment events in regimes with low star formation rates has been shown to be important for setting the width of stellar abundance patterns in low metallicity environments of the Milky Way halo \citep[e.g.][]{Cescutti2008,Cescutti2014}, can play an even greater role in the physical evolution of these low mass dwarf galaxies \citep{Applebaum2018,Su2018} and their abundances \citep[e.g.][]{RomanoStarkenburg2013,Romano2015,Ojima2018}. Indeed the increase in scatter in the stellar abundance patterns of low mass dwarf galaxies and UFDs has been attributed to inhomogeneous mixing and stochastic effects \citep[e.g.][]{Norris2010a,Lee2013,Simon2015,Mashonkina2017,Suda2017}. While chemical evolution models can account for some of these effects, they are largely unable to account for the detailed hydrodynamics interactions and turbulent mixing in a multi-phase ISM that may enhance (or smooth over) these stochastic effects. This complicates the interpretation of abundance patterns in these galaxies, particularly in attempting to assign a single (or small number) of enrichment events to specific abundance features \citep[e.g.][]{Ji2015,Fraser2017,Chen2017,Ishigaki2018,Welsh2019,Hartwig2019}

Recent cosmological zoom simulations have attained sufficient resolution to follow the evolution of individual low mass dwarf galaxies and UFDs (or their progenitors at high redshift) and can investigate their chemical properties \citep{Jeon2017,Maccio2017,Christensen2018,Corlies2018,Escala2018,Munshi2018,RevazJablonka2018,wheeler2019,Agertz2019}. 
In particular, \cite{Cote2018} compares high-resolution hydrodynamics simulations directly with one-zone models, finding the importance of non-uniform mixing in driving abundance spreads in these galaxies, and characterize multiple hydrodynamics effects that are challenging to parameterize in current one-zone models. Additional works have conducted more direct investigations into what drives the enrichment process for individual sources using both hydrodynamics simulations \citep{PanScannapiecoScalo2013,Ritter2015,Safarzadeh2017,Hirai2015,Hirai2017,Emerick2018b,HaynesKobayashi2019} and semi-analytic models \citep[e.g.][]{KrumholzTing2018}. 
In spite of this progress, there is still substantial work to be done in understanding the physical processes that drive evolution of both the mean and width of stellar abundances in low-mass dwarf galaxies.

The astrophysical origin of $r$-process enrichment is still highly uncertain (see \citet{Thielmann2017}, \citet{Frebel2018}, \citet{Cote2019}, and \citet{Cowan2019} and references therein) with possible origins including core collapse supernovae, binary neutron star (NS) mergers, NS-black hole mergers, magneto-rotational supernovae (SNe) \citep[e.g.][]{Winteler2012}, jet-driven SNe, and collapsars \citep[e.g.][]{Siegel2019} . Metal poor stars in the Milky Way's halo and nearby dwarf galaxies provide some of the greatest sensitivity to potential $r$-process sources. The $r$-process origin in these environments has has been investigated in analytic or semi-analytic models \citep[e.g.][]{Beniamini2018,Macias2018,Macias2019,SchonrichWeinberg2019,Wehmeyer2019} and directly in cosmological hydrodynamics simulations which either directly included models of $r$-process enrichment or placed these enrichment events in halos by-hand \citep[e.g.][]{Shen2015,vandeVoort2015,vandeVoort2019,Safarzadeh2017}. Broadly, these works generally find a preferred source of $r$-process enrichment to best match observations, but no model has yet been able to reproduce all observed stellar abundance trends across environments. However, as discussed earlier, non-uniform mixing plays an important role in galactic chemical evolution and its significance in these environments (the Milky Way halo and UFDs) may dramatically influence how one should use abundance patterns to constrain $r$-process yields. In addition, if much of these metals are ejected from low mass UFDs through galactic winds, this will change estimates for the total mass of $r$-process yields implied by observed stellar abundances. Finally, if metal mixing and ejection properties vary significantly between different sources (e.g. NS-NS merger and collapsars), this may provide an additional important discriminator between potential sites of $r$-process enrichment. Investigating these effects in detail requires simulations capable of capturing individual enrichment events with distinct injection energies, as compared to models utilizing smoothed enrichment from simple stellar populations. 

In \cite{Emerick2018b} we examined metal mixing in simulations of an isolated, low-mass dwarf galaxy following stellar feedback and chemical enrichment on a star-by-star basis. By following individual enrichment sources---asymptotic giant branch (AGB) winds, massive stellar winds, core collapse SNe, and Type Ia SNe---we were able to resolve differences in how metals from these sources evolve in a low mass dwarf galaxy. We found that elements released through AGB winds (e.g. s-process elements) have broader abundance distributions in the ISM than elements released in SNe (e.g. $\alpha$ and Fe-peak elements). In addition, AGB wind elements coupled more weakly to the significant galactic winds from this dwarf galaxy, and were retained at a much higher fraction ($\sim 20\%$) than elements from SNe ($\sim 5\%$). However, this is the result from many enrichment sources over an extended period of time in an isolated dwarf galaxy. It is unclear how much metal mixing varies across individual sources. 

In this work we utilize the detailed simulations introduced in \cite{Emerick2019} to conduct a controlled set of ``mixing experiments'' whereby we restart each simulation with enrichment events placed by hand in order to more directly investigate the evolution of metals from an individual event. While elemental yields were tracked for each star in our simulation, we lacked the necessary Lagrangian information\footnote{The history of a given mass element---which is followed natively in smoothed particle hydrodynamics and other Lagrangian codes---but not in the Eulerian grid-based hydrodynamics simulations used here.} about the metals once they were released into the ISM to be able to trace the evolution of single enrichment events. We investigate primarily how the feedback ejection energy of individual sources $E_{\rm ej}$ and global star formation rate (SFR) at the time of enrichment affect how metals are ejected from the galaxy in galactic winds and mix into the ISM. In Section~\ref{sec:methods} we briefly outline our methods and discuss the setup of these mixing experiments. In Section~\ref{sec:results} we discuss the role that $E_{\rm ej}$, global SFR, radial position of the enrichment event, and local ISM density around each enrichment event affects metal abundance evolution. We discuss these results and conclude in Section~\ref{sec:discussion conclusion}.

\section{Methods}
\label{sec:methods}
We refer the reader to \cite{Emerick2019} for a detailed description of our numerical methods, initial conditions, feedback, and chemical evolution model. We briefly summarize the key components of these methods below.

This work follows the evolution of an idealized, isolated, low-mass dwarf galaxy with an initial gas mass of $M_{\rm gas} = 1.80 \times 10^6$~M$_{\odot}$ initialized as an exponential disk with radial and vertical scale heights of 250~pc and 100~pc respectively. This galaxy is embedded in a static, \cite{Burkert1995} dark matter potential with virial mass and radius $M_{\rm vir} = 2.48\times 10^{9}~M_{\odot}$ and $R_{\rm vir}~=~27.4$~kpc. This is evolved using the adaptive mesh refinement hydrodynamics code \textsc{Enzo} \citep{Enzo2014}, with a minimum/maximum spatial resolution of 921.6~pc / 1.8~pc. The grid is refined to maintain a mass resolution of 50~M$_{\odot}$ per cell, and to ensure that the Jeans length is resolved by at least eight cells. In addition, a three-zone radius region around any star particle that has active feedback (stellar winds or SNe) is refined to the maximum grid resolution. We use the chemistry and cooling package \textsc{Grackle} \citep{GrackleMethod} to solve a nine species non-equillibrium chemistry model that includes gas-phase and dust H$_2$ formation, a uniform UV background, and localized self-shielding. 

\subsection{Star Formation and Stellar Feedback}
Our simulation stochastically forms star particles in dense gas ($n > 200$~cm$^{-3}$) by randomly sampling a \cite{Salpeter1955} IMF and depositing individual star particles from 1~M$_{\odot}$ to 100~M$_{\odot}$. For stars above 8~M$_{\odot}$, we follow their H~{\sc i} and He~{\sc i} ionizing radiation using the adaptive ray-tracing radiative transfer method of \cite{WiseAbel2011}, and trace their radiation in the Lyman-Werner and FUV bands using an optically thin approximation. These stars eject mass and energy over their lifetimes through stellar winds, and we include mass and thermal energy injection of both core collapse and Type Ia SNe. Stars below 8~M$_{\odot}$ have no feedback during their lifetime, except mass and energy deposition of their AGB winds at the end of their life. For stellar winds and SNe, mass, energy, and metals are injected to the grid by mapping a three-cell spherical region ($r=3 \times dx = 7.2~$pc) to the grid using a cloud-in-cell interpolation scheme. 

\subsection{Mixing Experiment Setup}
\label{sec:experiment}
We restart the fiducial, full-physics simulation described in \cite{Emerick2019} at two different times: 180 Myr, and 360 Myr, labeled as runs \runonenu~and \runtwonu~respectively. These correspond to two different points in the galaxy's star formation history, testing how much variance with the SFR is expected in the metal mixing and ejection. Run \runonenu~occurs during the lull in star formation ($\dot{M}_{*} \sim 6 \times 10^{-5}$~M$_{\odot}$~yr$^{-1}$) following the initial SFR peak ($\dot{M}_{*} \sim 10^{-3}$~M$_{\odot}$~yr$^{-1}$), and run \runtwonu~occurs in an extended period of little to no ongoing star formation ($\dot{M}_{*} < 1 \times 10^{-5}$~M$_{\odot}$~yr$^{-1}$). We attempted to to evolve each simulation for 150 Myr, but due to computational constraints this was not always possible.


\begin{deluxetable}{cccc}

\tablecaption{Energy and mass injected for each type of source tested. Energy is injected as pure thermal energy to the simulation grid, but we compute the corresponding ejection velocities for comparison to the escape velocity (see text for more detail).
\label{table:setup}}
\tablehead{
\colhead{Energy (erg)} & \colhead{$m_{\rm ej}$ (M$_{\odot}$)} & \colhead{$v_{\rm ej}$ (10$^{3}$ km s$^{-1}$)} \\
}
\startdata
$10^{46}$ & 2.0  & $22.$ \\
$10^{49}$ & 0.01 & $10.$ \\
$10^{50}$ & 0.01 & $32$ \\
$10^{51}$ & 10.0 & $0.31$ \\
$10^{52}$ & 2.5  & $22.$ \\
\enddata
\end{deluxetable}

At the beginning of each restart, we place by hand one or more enrichment events at assigned positions throughout the galaxy, with thermal injection energies ($E_{\rm ej}$) and masses ($m_{\rm ej}$) as given in Table~\ref{table:setup}. We compute the corresponding ejection velocity assuming injection into a vacuum (ignoring swept-up ISM mass, which can be significant) as $v_{\rm ej} = \sqrt{2 E_{\rm ej}/m_{\rm ej}}$. The escape velocity throughout the galaxy is nearly constant at about 81~km~s$^{-1}$. We vary $E_{\rm ej}$ and $m_{\rm ej}$ to sample the range of ejection energies associated with significant sources of chemical enrichment,  including AGB winds ($10^{46}$~erg; $\sim 2$ M$_{\odot}$), NS-NS mergers ($10^{49} - 10^{50}$~erg; $\sim 0.01$ M$_{\odot}$), SNe (10$^{51}$~erg; $\sim 10$ M$_{\odot}$), and exotic enrichment sources, such as hypernovae (HNe), that can reach much higher energies (10$^{52}$~erg; $\sim 10$ M$_{\odot}$). \footnote{We note that---due to an error in the setup of these runs---$m_{\rm ej}$ for the HNe-like event (10$^{52}$~erg) is low by a factor of four. We do not expect this to significantly impact the conclusions of this paper.} The injection masses are intended to be order of magnitude estimates of corresponding astrophysical sources, but we note that they are ultimately somewhat arbitrary as they are often small compared to total ISM mass in the 3-zone radius injection region, let alone in the swept-up ISM outside this region (for reference, the total mass in an injection region for $n_{\rm ISM} \sim 1$~cm$^{-3}$ is about 20~M${\odot}$).

%

Each event deposits mass into a corresponding passive scalar tracer field---unique to that event---to trace how the metals from these sources mix in the ISM over time. Each run contains only sources from a single event type, as indicated in the run-name by the log of the injection energy in ergs. For example, the run beginning at 180~Myr with AGB-like events is labeled ``\runone E46". We place multiple events per run, spread over the galaxy to test how radial and azimuthal position in the galaxy affects mixing and ejection, but limited to ensure that the events do not overlap and influence each other dynamically. For the low-energy events, we are able to use 19 events per run, while the $10^{49}-10^{51}$~erg runs contain 7 events, and the 10$^{52}$~erg runs only contain a single event.

\section{Results}
\label{sec:results}
Perhaps the four most important parameters to quantify for each enrichment event are: 1) what fraction of released metals are immediately available for star formation, 2) how does this fraction evolve over time as metals cool from hot phases into star forming gas, 3) what fraction of metals are carried out of the galaxy in outflows, and 4) how homogeneously distributed are the released metals as a function of time. We discuss the first three points in Section~\ref{sec:ISM CGM}, focusing on the average behavior of multiple enrichment events at fixed ejection energy and address the final quantity in Section~\ref{sec:spreads} and Section~\ref{sec:inhomogeneity}. Finally, in Section~\ref{sec:individual event variation} we discuss how these properties vary with individual events, and, in Sections~\ref{sec:radial position} through ~\ref{sec:SFR}, how they vary with radial position in the galaxy, ISM properties in the event region, and global galaxy SFR.

For simplicity, throughout this work we consider the ISM of our dwarf galaxy to be all gas within a fixed cylindrical region of radius 600~pc and $|z| < 200$~pc centered on the galaxy. The circumgalactic medium (CGM) is all gas outside of this disk, but within the virial radius (27.4~kpc). We split the ISM into four phases: cold neutral medium (CNM; $T < 10^2$~K), warm neutral medium (WNM; $10^{2}~\rm{K} \leq T < 10^{4}~ \rm{K}$), warm ionized medium (WIM; $10^{4}~\rm{K} \leq T < 10^{5.5}~\rm{K} $), and hot ionized medium (HIM; $T \geq 10^{5.5}$~K). Unless otherwise specified, when we refer to metals ejected from our dwarf galaxy we mean metals that are no longer within the ISM, and are either within the CGM or have been ejected beyond the virial radius.

\subsection{Enrichment of the ISM and CGM}
\label{sec:ISM CGM}

\begin{figure*}
\centering
\includegraphics[width=0.45\linewidth]{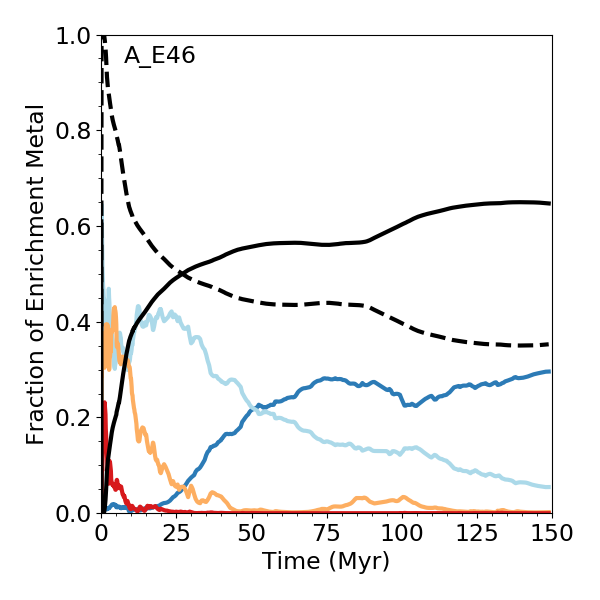}
\includegraphics[width=0.45\linewidth]{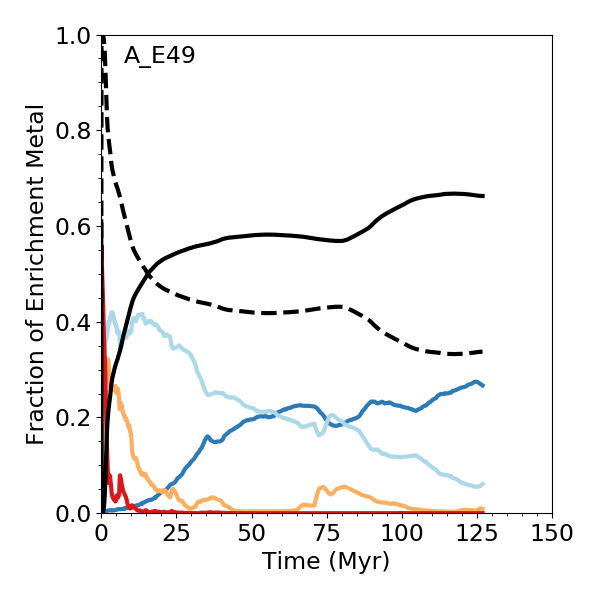} \\
\includegraphics[width=0.45\linewidth]{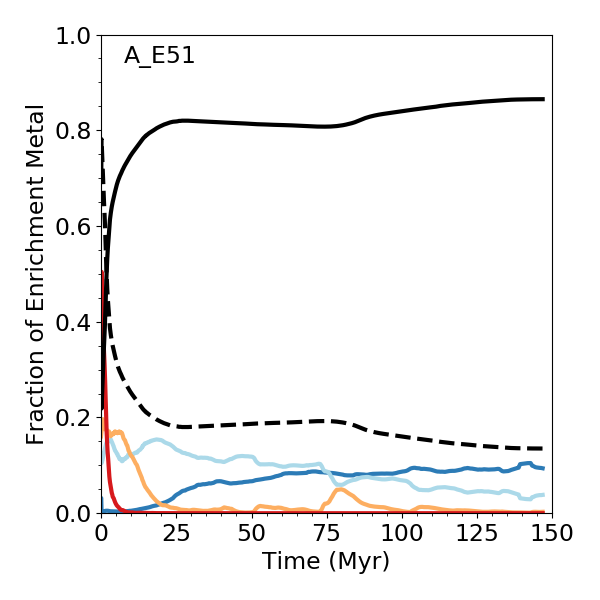}
\includegraphics[width=0.45\linewidth]{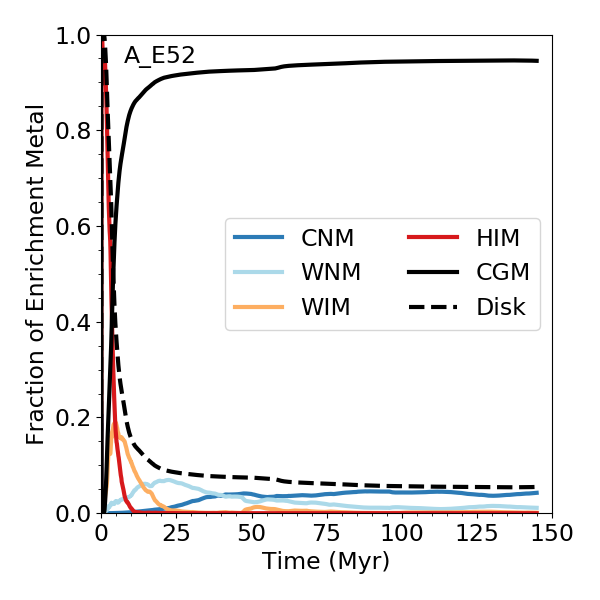}
\caption{Time evolution of the fraction of metals in each phase of the ISM (colored lines; CNM: dark blue, WNM: light blue, WIM: orange, HIM: red), the galaxy's disk (black, dashed), and the CGM (black, solid) as averaged across all events at a given $E_{\rm ej}$. The fractions are all normalized to the total amount of metals initially injected in each event. The individual ISM phases sum to the black, dashed line.}
\label{fig:ISM_CGM}
\end{figure*}

\begin{figure*}
  \centering
  \includegraphics[width=0.45\linewidth]{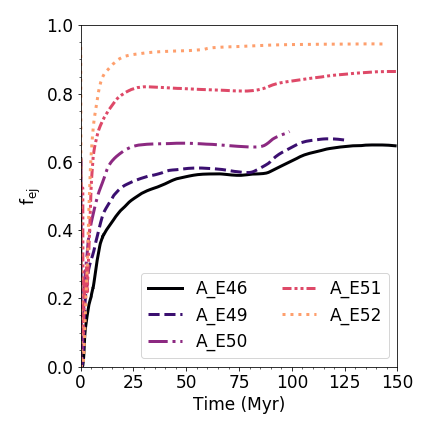}
  \includegraphics[width=0.45\linewidth]{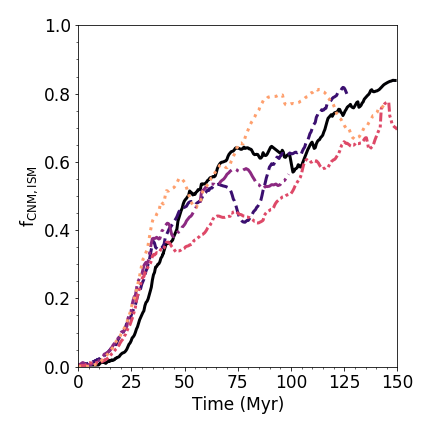}
  \caption{Time evolution of the fraction of enrichment source tracer metals ejected from the galaxy (left), and the fraction of tracer metals within the ISM that are in the CNM (right). These show the same data as the lines in Figure~\ref{fig:ISM_CGM}, but the CNM lines are now normalized by the black, dashed total ISM line in Figure~\ref{fig:ISM_CGM}.  }
  \label{fig:CGM_CNM}
\end{figure*}

We summarize the source-averaged results for all types of enrichment events in Figure~\ref{fig:ISM_CGM} by showing the fraction of source metals contained within each phase of the ISM (colored lines, which sum to the black dashed line), and the CGM (black, solid). These plots show a clear, immediate trend across $E_{\rm ej}$ for all phases shown in the figure, as well as for the distribution between ISM (black, dashed) and CGM (black, solid) lines. In \cite{Emerick2018b}
we demonstrated that there was a significant difference in metal ejection fraction, $f_{\rm ej}$, for metals from AGB sources as compared to metals released in SNe. Here, we confirm that this is driven by differences in the energy of the events. Figure~\ref{fig:ISM_CGM} shows that material ejected in events with higher $E_{\rm ej}$ is much more readily ejected from the disk of the galaxy than material from lower $E_{\rm ej}$ events. 
The high-energy events rapidly converge on a peak $f_{\rm ej}$ within about 20~Myr of the event, with only a gradual increase towards the end of the 150~Myr as metals in the ISM are swept up in additional outflows. The lower energy events evolve more gradually. The lowest energy event, corresponding to AGB winds, reaches $f_{\rm ej} \sim$~0.68 by the end of the simulation, lower than the 0.87 for the 10$^{51}$~erg events, and 0.95 for the 10$^{52}$~erg event. To emphasize these differences, we plot all the CGM fractions together
in the left panel of Figure~\ref{fig:CGM_CNM}.

In all cases, the tracer metals are initially deposited predominantly in the ionized phases of the ISM, the WIM or the HIM, with the relative fraction in each phase driven by the energy of the event. Metals in the ISM injected with $E_{\rm ej} > 10^{49}$~erg are initially located predominantly in the HIM, roughly half for $E_{\rm ej} = 10^{49}$~erg and nearly all for $E_{\rm ej} > 10^{51}$~erg. The lowest energy events are initially in the WIM and WNM, tracing the two dominant volume-filling components of the ISM. These events do not have sufficient energy to generate HIM by themselves, while the E46 events can only ionize gas at lower densities (n~$\lesssim 0.1$~cm$^{-3}$).

Gas above the star formation threshold in our simulations is limited at any one snapshot, and short-lived. As a proxy, we examine the evolution of the CNM, from which the star forming gas originates. In the right panel of Figure~\ref{fig:CGM_CNM}, we examine the evolution of the metal fraction of the CNM \textit{for just those elements retained in the ISM}. In general, very few of these elements are available for immediate star formation in the CNM ($<< 1\%$, see Figure~\ref{fig:CGM_CNM}). Although the initial CNM fractions are about the same for each source, the evolution over the first $\sim$50~Myr is qualitatively different.  The metals from higher energy sources with $E_{\rm ej} > 10^{49}$~erg are more rapidly incorporated into the CNM than the metals in E46, even though the former models retain a lower fraction of metals in the ISM. This is most significant at $\sim$20~Myr, when the fraction of metals in the CNM from these sources is a factor of $\sim$3--4 higher than the E46 metals. By $\sim$50~Myr, the fractions become similar, with no clear trend as a function of $E_{\rm ej}$. Thus, metals from the lowest energy sources (i.e. AGB stars) only start to become available for enrichment of future sites of star formation $\sim$20 Myr after metals from higher energy sources.

\cite{SchonrichWeinberg2019} find that models of $r$-process enrichment in the Milky Way can better fit observations provided the fraction of metals immediately available for star formation from NS-NS mergers is greater than that of SNe. Based on our simulation, we do not find evidence that this fraction is different immediately after each event, nor do we find a clear trend between longer term differences and injection energy for sources with 10$^{49}$-10$^{51}$~erg. However, we note that we may not have sufficient resolution to properly resolve the details of the initial mixing of individual enrichment events in the ISM. In addition, this value will be sensitive to whether or not a given event is more (or less) likely to occur in the vicinity of or inside an active, star forming region, or far from dense gas in the ISM. We are also missing important physical processes, such as dust production in AGB winds and core collapse SNe and differences in cooling rates with the abundances of individual elements, which may change how rapidly elements from a given source are incorporated into the cold ISM. Investigating the fraction of metals immediately available for star formation requires even higher resolution simulations of metal mixing, such as those that following mixing in and around individual star forming regions \citep[e.g.][]{Kuffmeier2016,Armillotta2018}.

The metals from each source do gradually tend towards a fraction of $\sim$0.8 by the end of the 150~Myr simulation time, which is approximately the total mass fraction of the CNM. This trend---that the fraction of metals contained in a given phase tends towards the mass fraction of that phase---is true across all phases in the simulation. Therefore, we can conclude that the metals for each source are well-mixed across the \textit{phases} of the ISM on timescales of $\sim$100--150~Myr over the whole galaxy. This is comparable to, but less than, the dynamical timescale of the galaxy, $\sim$200~Myr. Although well-mixed across phases, we emphasize that this does not imply that the abundances are the same across phases nor that the metals are spatially well-mixed across the galaxy. We investigate this second point further below.

\subsection{Homogeneity of Mixing}
\label{sec:spreads}

\begin{figure*}
\centering
\includegraphics[width=0.9\linewidth]{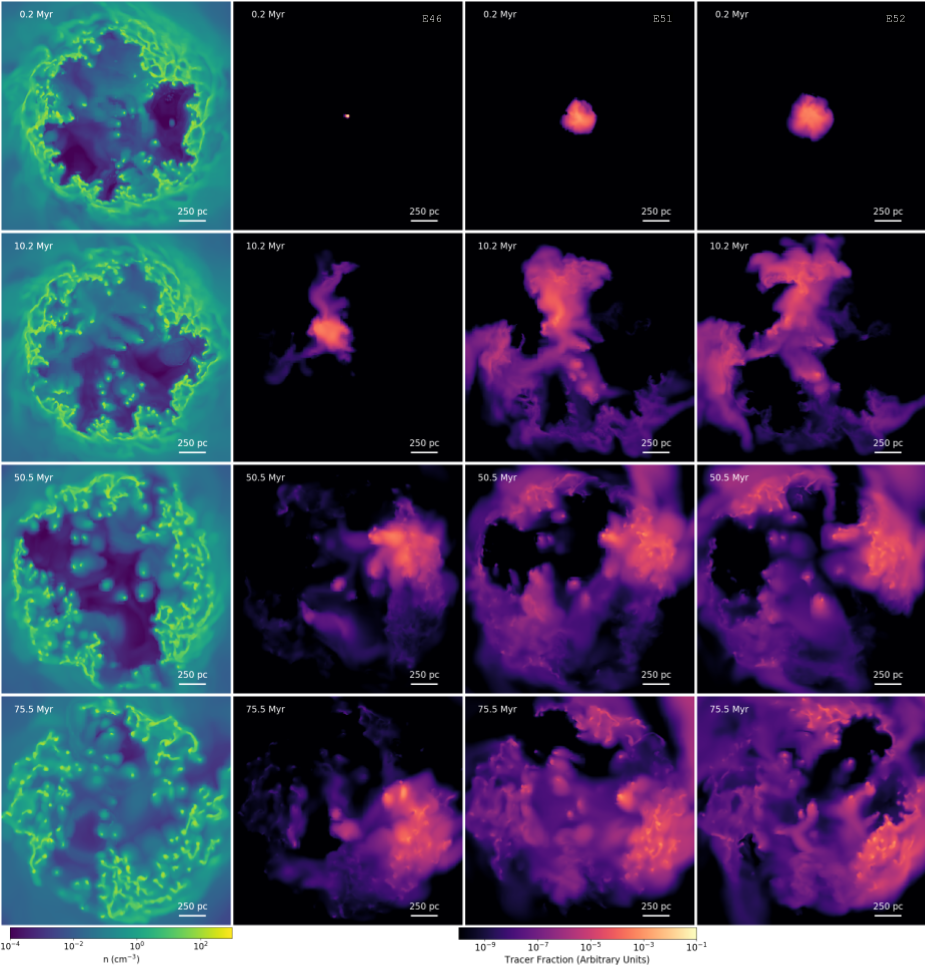}
\caption{Face-on evolution of the enrichment from sources placed at the center of the galaxy. We show a projection of the gas number density in the ISM (far left) and the enrichment evolution of an E46, E51, and E52 event at times after the initial injection of 0.2 Myr, 10 Myr, 50 Myr, and 75 Myr. The enrichment panels show the total mass of the tracer metal in each pixel. The right three panels have no weighting except a uniform normalization to maintain the same color scaling across the three injection sources. Since the initial injection mass of each event varies and the amount ejected / retained by the ISM varies, each panel is normalized by both the mass of tracer metal initially injected and the mass of that metal retained in the ISM.}
\label{fig:mixing_panel}
\end{figure*}

\begin{figure*}
\centering
\includegraphics[width=0.9\linewidth]{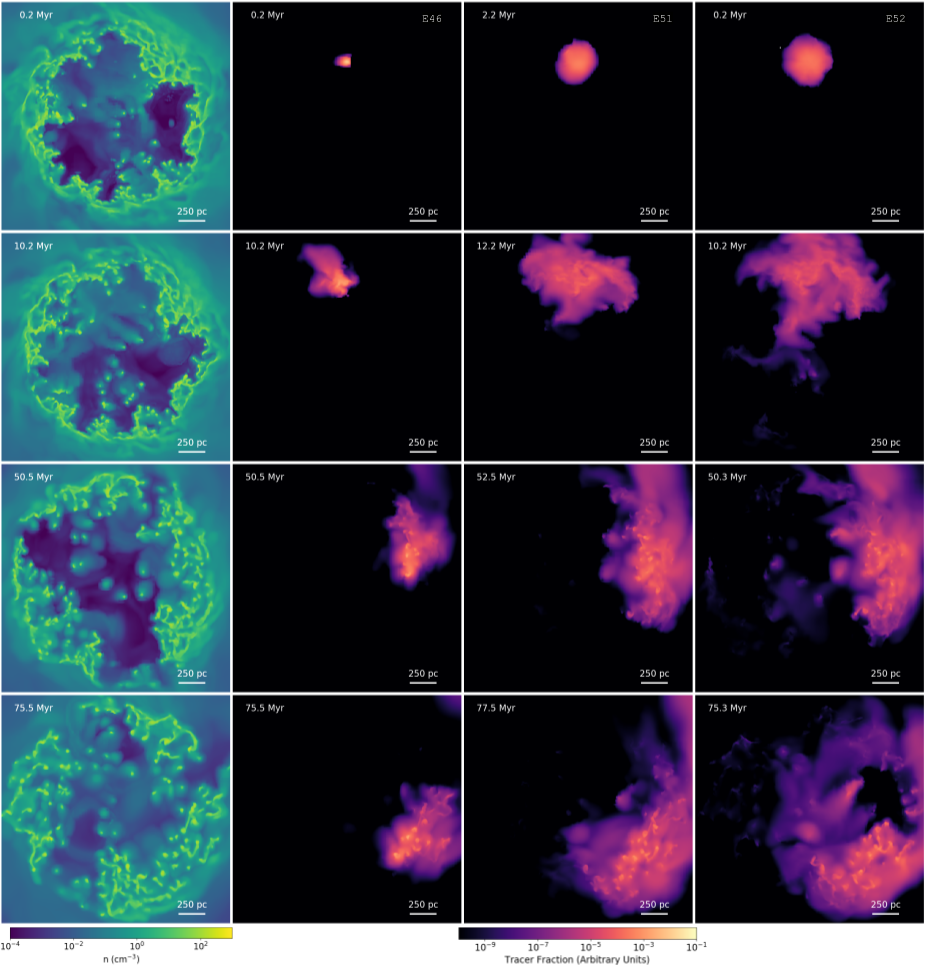}
\caption{The same as Figure~\ref{fig:mixing_panel}, but for enrichment events placed in the mid-plane, 300~pc from the center of the galaxy.}
\label{fig:mixing_panel2}
\end{figure*}

We show the spatial evolution of metals for different enrichment energies in Figures~\ref{fig:mixing_panel} and Figures~\ref{fig:mixing_panel2} to build intuition on how metals evolve in this galaxy over time. Each figure shows the time evolution of the abundance of a single tracer field (right three panels) associated with an enrichment event placed in the center of the galaxy (Figure~\ref{fig:mixing_panel}) and another event in the mid-plane, placed 300~pc from the center (Figure~\ref{fig:mixing_panel2}). The left panels show projections of the disk number density to illustrate the structure of the ISM in the galaxy over the four sampled points in time after the initial injection. The abundance panels show (left to right) the evolution of an E46, E51, and E52 source, corresponding to an AGB-like, SN-like, and HNe-like enrichment event.

These figures demonstrate that the injection energy of each enrichment event drives qualitative differences in where and how quickly the tracer metals are distributed over the galaxy. In both cases, the tracer metals from the lowest energy event (E46) take longer to evolve out of the initial injection region, and reach a smaller volume of the galaxy by the final panel. In the central injection event, the two higher energy sources distribute their metals over much of the galaxy within 10~Myr, enriching comparable (but not identical) regions by the final panel. Comparing the enrichment panels to the gas number density, there is a significant amount of tracer metals contained within dense clumps of gas.\footnote{As noted in the caption to Figure~\ref{fig:mixing_panel}, we emphasize that the images in the right three panels are not weighted. The color field is the total mass of the tracer metal in each pixel with a normalization.} However, although they contain much of the metals, these dense regions are considered metal poor; they have low metal mass fractions due to their significant total mass. While the total mass of tracer metals contained in the diffuse gas is less than that of these cold dense clumps, the diffuse gas is more highly enriched due to their higher metal mass fractions. Metals locked in these cold regions remain there, mixing poorly with the rest of the ISM, unless blown apart through star formation and stellar feedback. 

Comparing Figure~\ref{fig:mixing_panel} with Figure~\ref{fig:mixing_panel2}, the position of the enrichment source in the galaxy does make a noticeable difference in the evolution of the tracer metals for a fixed injection energy. In all cases, the tracer metals mix with a much smaller volume of the galaxy than their counterparts injected at the center of the galaxy. By the final panel, the tracer metals are still confined to about half (or less) of the galaxy. We better quantify the inhomogeneity of this mixing and how it varies with injection energy, event position, and galaxy properties below.

\subsection{Quantifying Inhomogeneity}
\label{sec:inhomogeneity}

To frame this analysis, let us first examine what processes affect the distribution of metals---the metal mass fraction PDF---in the ISM. The mean metal mass fraction for a collection of gas is simply the total mass in metals within that gas divided by the total gas mass. For a collection of $i$ distinct, homogeneous, parcels of gas, the mean metal mass fraction 
\begin{equation}Z_{\rm gas} =  \Sigma_i Z_iM_{{\rm gas},i} / \Sigma_i M_{{\rm gas},i}, \end{equation}
where $M_{{\rm gas},i}$ is the mass of the $i^{\rm th}$ parcel of gas, and $Z_{i}$ its metal mass fraction. A completely homogeneous ISM would have a $\delta$-function metal mass fraction PDF located at the mean metal fraction: $p(Z) = \delta(Z_{\rm gas})$. Metal mixing in real galaxies does not produce perfectly homogeneous distributions, but rather tends towards a Gaussian at long timescales \citep[e.g.][]{EswaranPope1988}; a "well-mixed" medium will still have non-zero spread. Here we would like to characterize how this spread evolves for single enrichment sources. In particular, we characterize how well the assumption of instantaneous, homogeneous mixing---commonly adopted in one-zone chemical evolution models---describes actual abundance evolution in a low mass dwarf galaxy.



\subsection{Abundance PDFs}
\label{sec:PDF}
\begin{figure*}
    \hspace*{-1.8cm}
  \includegraphics[width=1.2\linewidth]{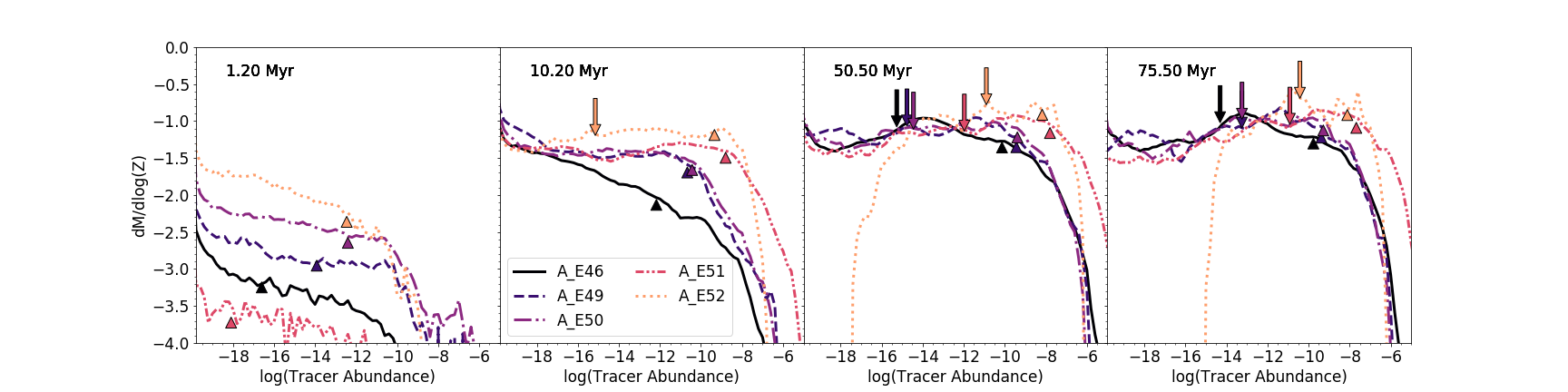}
     \hspace*{-1.8cm}
  \includegraphics[width=1.2\linewidth]{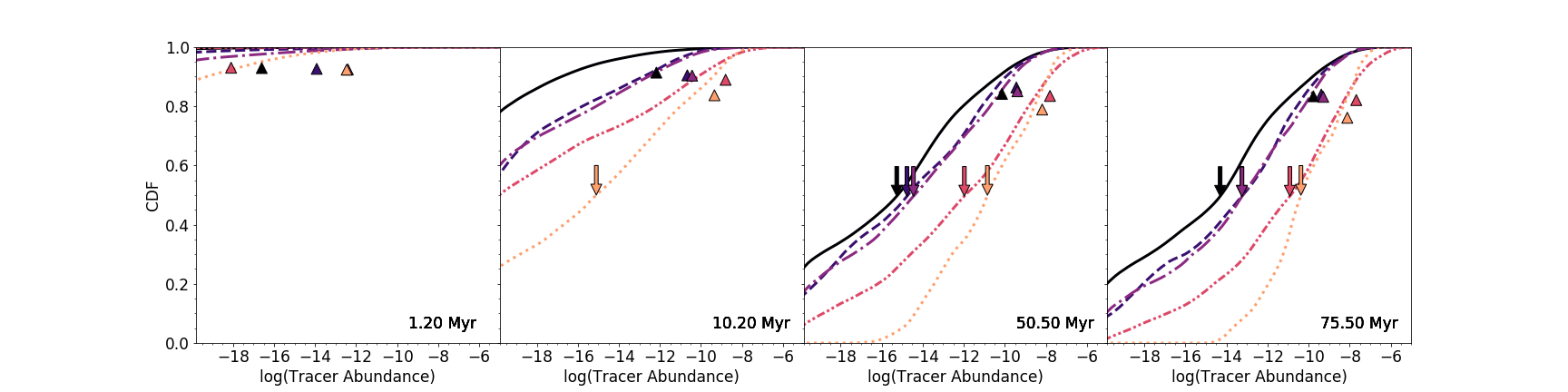}  
  \caption{The averaged metal mass fraction PDF (top) and CDF (bottom) for each enrichment event at times corresponding to the spatial distributions in Figure~\ref{fig:mixing_panel} and Figure~\ref{fig:mixing_panel2} (we use 1.2~Myr here instead of 0.2~Myr since the initial PDF is effectively a delta function at zero abundance). The median value of each PDF is marked by the downward pointing arrow, and the mean is given by the upward pointing triangle (and is to the right of the median in all cases). 
  Median arrows are not shown when less than half the gas has been enriched above the initial abundance (10$^{-20}$). For brevity, we plot only the distributions in the CNM (see text). In both, the bin spacing is 0.2~dex.}
  \label{fig:CNM-PDFS}
\end{figure*}

We begin by examining in Figure~\ref{fig:CNM-PDFS} the probability distribution functions (PDFs) and the corresponding cumulative distribution functions (CDFs) of the tracer metals at the same times as Figure~\ref{fig:mixing_panel} and Figure~\ref{fig:mixing_panel2}. We focus on the cold neutral medium---the source of star forming gas---as it dominates the total mass fraction in the disk of this galaxy. In each panel, the median mass fraction is denoted by the downwards pointing arrow, while the mean is denoted by the upwards triangle. Initially---in all cases---nearly all of the gas in the galaxy contains arbitrarily little tracer metal mass as it it confined to a relatively small volume. However, this rapidly evolves, forming a broad distribution over many orders of magnitude in tracer metal abundance with no clear distinct enriched and un-enriched components.\footnote{At early times the rate at which this occurs is controlled by the significance of numerical diffusion in our simulation. Therefore, this behavior depends on the exact numerical methods used to model with problem. While characterizing the numerical diffusion with an effective diffusion coefficient would aid in comparing to results across simulation methods, doing so in grid-based codes is possible only in very particular contexts \citep[see][]{deAvillez2002}.}

Three general trends can be seen in these distributions: 1) the mean abundance evolves minimally after 10~Myr, corresponding to the timescale at which $f_{\rm ej}$ plateaus (Figure~\ref{fig:CGM_CNM}), 2) the mean abundance is greater than the median at all times, and by many orders of magnitude, and 3) higher energy injection events evolve more rapidly towards higher abundances with narrower distributions. We quantify further the time evolution of the width of these distributions in the next section.

\subsection{Spread}
\label{sec:spread}

We use two quantities to characterize the spreads of these distributions. Both show the same general trends with energy (as discussed below), but serve two purposes.
The difference between the mean and median metal abundances represents the difference between abundances assuming perfect homogeneous mixing (mean) and the abundance of a typical collection of gas (median) in the ISM. Due to numerical diffusion, this difference in our simulations is a rough lower limit on the offset between typical gas-phase abundances in an inhomogeneous ISM and that assumed in a one-zone model.  While informative, comparing this quantity across simulations and with real galaxies is not straightforward. To aid in this comparison, we also compute the root-mean-square (rms) abundance deviation for only gas enriched above a fixed fraction, 10$^{-5}$, of the mean abundance at any given time. The exact choice of this value is somewhat arbitrary, but provides a consistent basis for comparison across works. As in Section~\ref{sec:PDF}, we focus on inhomogeneity in the CNM alone.

\begin{figure*}
  \centering
\includegraphics[width=0.475\linewidth]{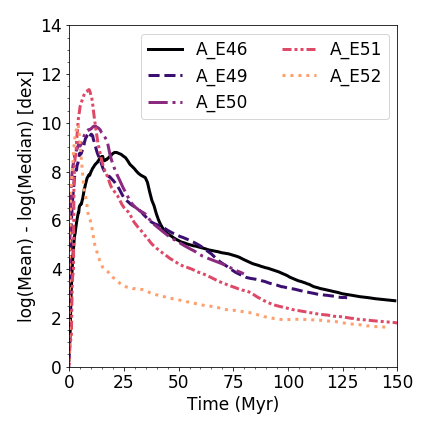}
\includegraphics[width=0.475\linewidth]{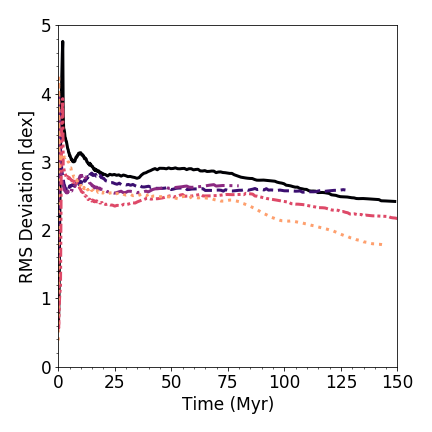}
\caption{The evolution of two measures of inhomogenity in the tracer abundances, the log(mean)-log(median) difference (in dex, left) and the rms deviation for all gas enriched to at least 10$^{-5}$ the mean abundance (right). At late times these two quantities are similar. Note the different vertical axis limits.}
  \label{fig:mean-median}
\end{figure*}

In Figure~\ref{fig:mean-median} we plot the time evolution of these measures of inhomogeneity for each source energy: the difference between log(mean) and log(median) (left, referred to as the mean-median difference for the remainder of this work), and the rms deviation of enriched gas (right). The evolution of the mean-median difference is easy to interpret conceptually. Enrichment in an initially pristine medium increases the mean-median difference by raising the mean metal fraction while keeping the median value fixed, so long as the newly enriched mass is a small fraction of the total gas mass. This is exactly the case for the enrichment experiments performed in this work. Conversely, preferential removal of metals from a medium will lower this quantity. Finally, the mixing of metal-rich gas elements with metal-poor will gradually bring the mean and median values to parity. The initial value for this spread is small in all cases, growing large (many dex) within the first 10~Myr. This delay is due to the different timescales over which most of the metals are incorporated into the cold gas, as shown in Figure~\ref{fig:ISM_CGM}. The large value of this spread is the result of the near-zero initial abundance of each tracer metal. Because of this, the exact magnitude of this initial spread is arbitrary, but it is its evolution---and how it compares across injection energies---that we are concerned with. The spread decreases significantly over the first 25~Myr or so, in part because metals are preferentially ejected from the ISM via outflows (lowering the average abundance, but leaving the median fixed) and in part due to mixing into the ambient ISM. 


As shown qualitatively in Figures~\ref{fig:mixing_panel} and~\ref{fig:mixing_panel2}, the injection energy for a given source leaves an indelible impact on the subsequent evolution of those metals in the ISM. Metals from \runone E46 are significantly less well-mixed than their higher-energy counterparts, \runone E51 and \runone E52. The evolution of this spread is qualitatively similar for the CNM. Across injection energies, there is an initial phase of more rapid mixing from the expansion of the initial injection event (most dramatic for \runone E52). Afterwards, mixing proceeds more slowly as the tracer metals diffuse throughout the galaxy via turbulent mixing in the ISM. The exact slope of this evolution is different for each event energy, and is more rapid for the higher energy events. Generally, these spreads appear to approach a plateau in their evolution, as most obvious in \runone E52. It is unclear if this value (about two dex in the CNM for \runone E52) will be the same for all energies, or at exactly what timescale it is reached for the lower energy events. 

\begin{figure}
  \centering
\includegraphics[width=0.95\linewidth]{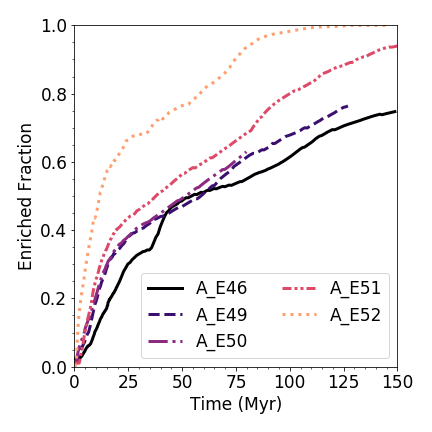}
\caption{The fraction (by mass) of the CNM that is enriched to within 10$^{-5}$ or more of the mean abundance for each enrichment source as a function of time.}
  \label{fig:enrichment_fraction}
\end{figure}

In the right hand panel of Figure~\ref{fig:mean-median} we isolate the inhomogeneity within enriched gas alone. Unlike the mean-median difference, this quantity exhibits significantly less evolution in each case. Ignoring the initial spike within the first Myr, the rms deviation decreases by---at most---a dex in each case. However, this quantity is still large, exceeding two dex. In general, the spreads here are ordered by energy, with higher energy events having a smaller spread, but the evolution of the lower energy events ($\leq$10$^{50}$~erg) are much closer to one another, and even intersect towards the end of the simulation time. At no point can any of these distributions be considered well-mixed within the 150~Myr simulation time, using either measure of inhomogeneity. Comparing these two measures of abundance spreads, it is clear that the different enrichment energies drive how much of the galaxy (by mass) is enriched to abundances near the mean tracer abundance. We confirm this in Figure~\ref{fig:enrichment_fraction}, which shows the enriched fraction---the mass fraction of gas in the CNM with abundance greater than 10$^{-5}$ times the mean abundance---for each source as a function of time. The enriched fraction is greater for higher energy enrichment events, driven predominantly by the initial growth of this quantity during the first $20-25$~Myr or so. After this point, the rate at which gas becomes enriched above 10$^{-5}$ times the mean is roughly the same across enrichment energies.

\begin{figure}
  \centering
  \includegraphics[width=0.975\linewidth]{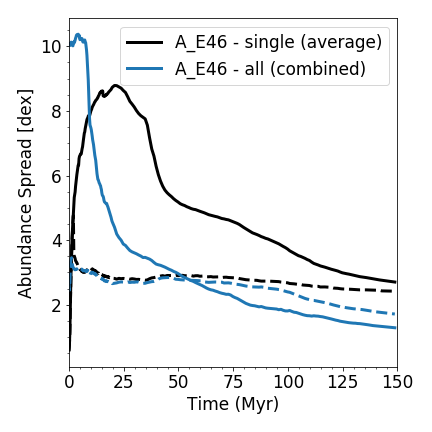}
  \caption{A comparison of the log(mean)-log(median) difference (solid lines) and the rms deviation (dashed lines) for the combined tracer metals of all 19 \runone E46 events (blue) to the typical behavior of a single event (the average of all 19, black). The combined field is computed by summing over the individual tracer fractions in each cell in the simulation, and thus represents the enrichment evolution of 19 identical, simultaneous events spread over the entire galaxy.}
  \label{fig:combined spread}
\end{figure}

We emphasize that these results show the behavior of metal enrichment from \textit{single} enrichment events over time. The mean-median / rms differences seen at the end of the simulation here are significant ($\sim$ 2--5 dex and $\sim$1.8--2.5 dex respectively). Our previous analysis of the total metal enrichment of all sources in this galaxy presented in \cite{Emerick2018b} found much smaller values: $\sim$0.2 dex for elements released in core collapse SNe and $\sim$0.5--0.8 dex for elements released in AGB winds. The only difference here is that the latter two values represent the abundance spreads obtained considering the ongoing contribution of many sources over the entire galaxy. As expected from prior works, this suggests that how common, how evenly, and how widely (spatially) distributed enrichment events are in a galaxy is an important determinant of abundance homogeneity. We demonstrate this point in Figure~\ref{fig:combined spread} by plotting the mean-median abundance difference for the total tracer metal fractions from all 19 \runone E46 events (see caption for more details), as compared to the averaged behavior as shown previously. The mean-median difference for the combined tracer (solid lines) is significantly lower than the typical value for a single source, dropping even below the spread seen in the highest energy single event. However, the rms deviation of enriched gas (dashed lines) is similar in both cases. The enrichment event frequency and distribution throughout the galaxy significantly affects the scatter in the associated abundances by changing the fraction of the ISM that is enriched. 

\subsection{Event-by-Event Variation}
\label{sec:individual event variation}

\begin{figure*}
   \centering
   \includegraphics[width=0.975\linewidth]{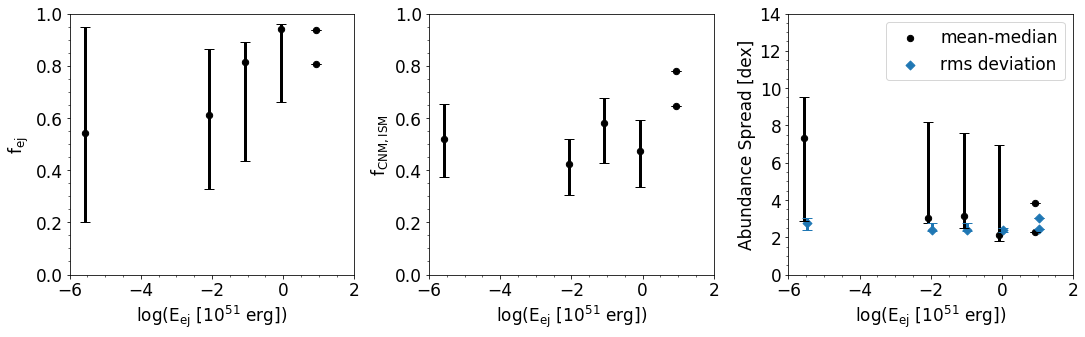}
   \caption{The variance in the fractions of metals ejected $f_{\rm ej}$ and retained in the CNM $f_{\rm CNM, ISM}$ (Fig.~\ref{fig:CGM_CNM}), and the mean-median separation and rms deviation of enriched gas (Fig.~\ref{fig:mean-median}). The rms deviation of enriched gas is plotted with a slight horizontal offset for clarity. There is significant variation in each quantity at fixed injection energy, with the exception of the rms deviation of enriched gas. We give the median values for each quantity at each energy (squares) at 75~Myr after the injection event, while the error bars represent the IQR at each energy. We show the results from \runone E52\_r0 and \runone E52\_r300 as individual points and do not estimate their IQR.}
   \label{fig:CGM_CNM_variance}
 \end{figure*}

For clarity we have focused on the \textit{mean} evolution of individual enrichment events at fixed injection energy. We turn now to discuss how much variety exists among these individual events. For each run from set \runonenu, Figure~\ref{fig:CGM_CNM_variance} shows the fraction of event tracer metals ejected from the galaxy (left), the fraction of metals in the ISM within the CNM (center), and the mean-median difference (right, black) and the rms deviation of enriched gas (right, blue) 75 Myr after each enrichment event. The median values at each energy are shown as points, while the error bars denote interquartile range (IQR). With the exception of the rms deviation of enriched gas, there is substantial variation at each energy for these quantities, with greater variation at lower energies. The similar spreads in enriched gas for different enrichment events suggests that abundance spreads in the ISM---once enriched---are driven predominantly by global galactic properties and not the details of the enrichment itself. Although there is a general trend in each of these quantities with injection energy, it is clearly not the only determining factor in the evolution of the tracer metals from each source. We investigate the dependence on a few of these properties below.

\subsection{Dependence on Radial Position}
\label{sec:radial position}
\begin{figure*}
   \centering
   \includegraphics[width=0.95\linewidth]{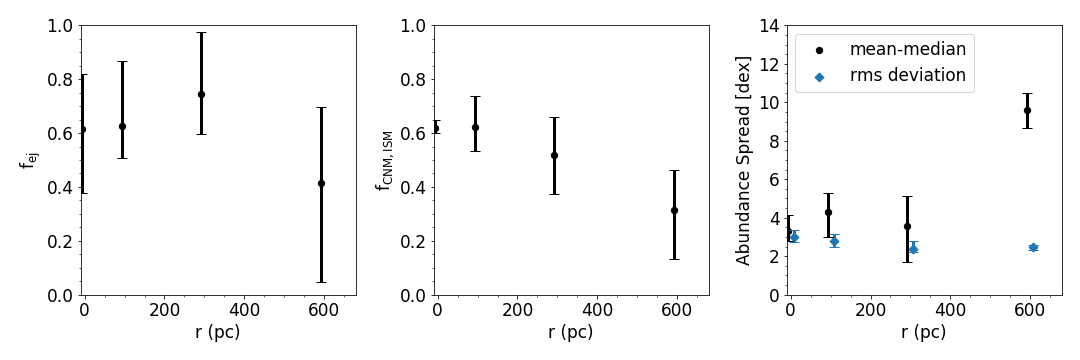}
   \caption{The dependence of $f_{\rm ej}$, $f_{\rm CNM, ISM}$, and the abundance spread on the radial position of each event at 75~Myr after injection as averaged over injection energy. As in Figure~\ref{fig:CGM_CNM_variance}, the points give the median at each radius and the error bars show the IQR.}
   \label{fig:radial position}
 \end{figure*}

We place each source at regular, but arbitrary, positions in the galaxy without consideration for the local ISM conditions at injection. Sources are placed at the center of the galaxy ($r = 0$~pc) or at various radii (at cardinal positions at 300~pc, and 600~pc), all in the mid-plane of the galaxy. Since the E46 events were likely to not self-interact, we additionally placed events at $r = 100$~pc and one event at each $r$ at about one scale height above the disk ($z = 50$~pc). For each \runonenu~run, we plot $f_{\rm ej}$, $f_{\rm CNM, ISM}$, and the mean-median difference as a function of radial position in the galaxy for each event---averaging over energy---in Figure~\ref{fig:radial position}. We do not consider the events above the mid-plane here since they sampled only a single injection energy.


The dependence of each of these quantities on the radial position is interesting. The ejection fraction $f_{\rm ej}$ increases with radius until the outermost edge, with significant spreads at each position. Star formation and the cold gas distribution in this galaxy is not uniformly distributed throughout the evolution, as can be seen in the number density panels of Figure~\ref{fig:mixing_panel}. Most of the star formation and feedback during this period occurs off-center from the galaxy, closer to the inner edge of the ring of dense clumps, and little to no star formation occurs at the outer edge. The increase of $f_{\rm ej}$ at intermediate radii could be explained by stronger feedback and outflows at these points that drive out metals. With this picture in mind, it is interesting that f$_{\rm CNM, ISM}$ decreases with radius, even though the outer region of the galaxy contains the most cold gas. This behavior stems from the E50 and E51 events (E46 and E49 show flat trends in median $f_{\rm CNM, ISM}$ with radius). 
Finally, the right panel shows that the galaxy-wide mean-median difference is not well correlated with position, with the exception of the farthest radius which is significantly less well mixed than events towards the inner galaxy. As in Figure~\ref{fig:CGM_CNM_variance}, the rms deviation of enriched gas does not show significant variations at fixed radii, but does show a very slight trend, decreasing with radius. 

\subsection{Variation with Local ISM Conditions}
\label{sec:ISM density}

\begin{figure*}
   \centering
   \includegraphics[width=0.95\linewidth]{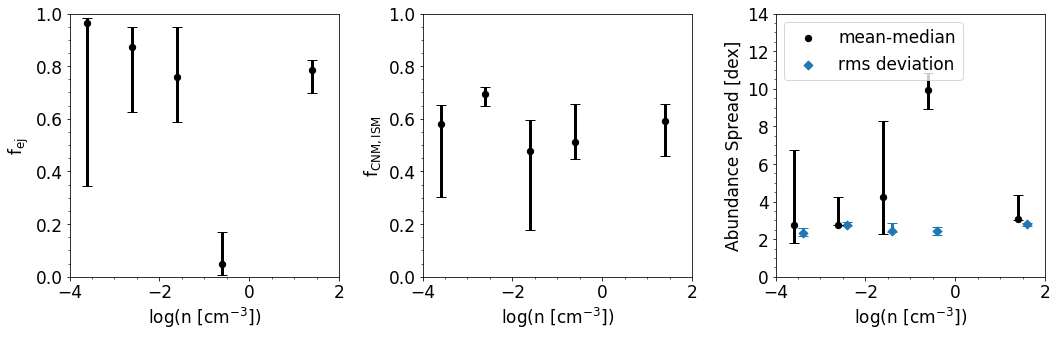}
   \caption{The variation of $f_{\rm ej}$, $f_{\rm CNM, ISM}$, and the spread of the metal fraction PDFs in the ISM as a function of the average local ISM number density ($n$) within the injection region of the event just prior (within 0.1~Myr) of the event. Results are binned using 1~dex bins in $n$. Points show the median value in each $n$ bin and error bars show the IQR (same as Figure~\ref{fig:CGM_CNM_variance}.)}
   \label{fig:ISM_variance}
 \end{figure*}

We compute the average ISM properties within a four-zone radius ($4~dx $~=~7.2~pc) around each injection site to examine any potential correlations in evolution with ISM properties. This fully envelopes the injection region for each event (a three-cell radius sphere mapped onto the grid with a cloud-in-cell interpolation scheme). In Figure~\ref{fig:ISM_variance} we plot $f_{\rm ej}$, $f_{\rm CNM, ISM}$, and the mean-median difference as a function of the average number density $\langle n \rangle$ in this region within 0.1~Myr (our time resolution) of the event. The ejection fraction $f_{\rm ej}$ is typically highest at low densities ($\log(n ~[\rm{cm}^{-3}]) < -1$). Metals ejected into this phase for higher energy events are less likely to be trapped by colder, dense gas, and more likely to be swept up in outflows from SN feedback; this value decreases significantly just above $\log(n ~[\rm{cm}^{-3}]) = -1$, and spikes again at the highest densities. At these high densities, gas is more likely to be star forming; feedback from newly formed stars readily removes the tracer metals from the galaxy. Interestingly $f_{\rm CNM,ISM}$ appears to be relatively independent of ISM density. The same processes that drive the trend in $f_{\rm ej}$ can explain the observed trend in the mean-median difference. Metals are more easily mixed when injected into low density gas, which allows for a larger initial expansion of the enrichment event and better coupling to the turbulent motions from stellar feedback; this is except for the highest density which, again, is affected by nearby star formation.


\subsection{Variation with SFR}
\label{sec:SFR}
\begin{figure*}
  \centering
  \includegraphics[width=0.45\linewidth]{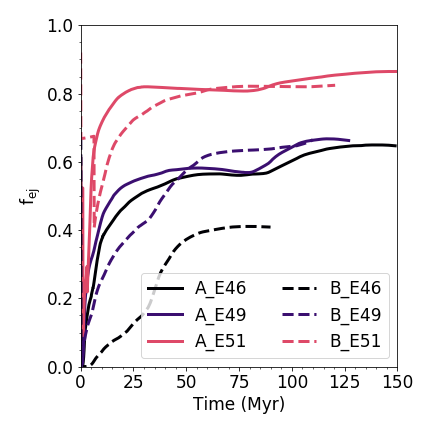}
  \includegraphics[width=0.45\linewidth]{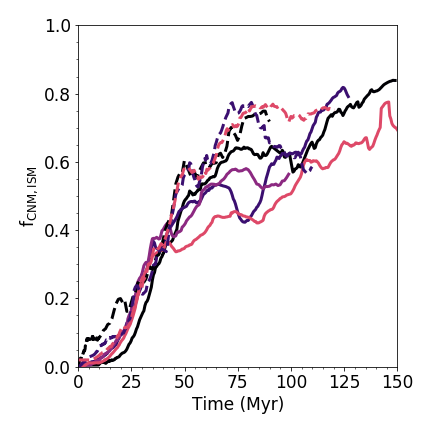}\\
  \includegraphics[width=0.45\linewidth]{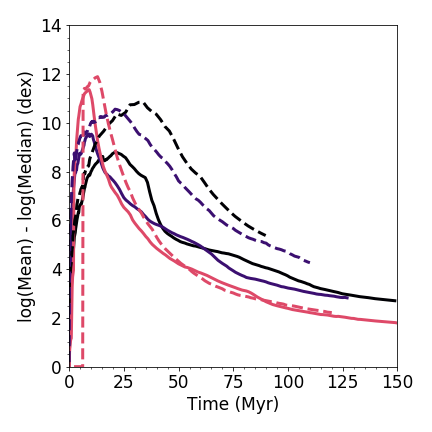}
   \includegraphics[width=0.45\linewidth]{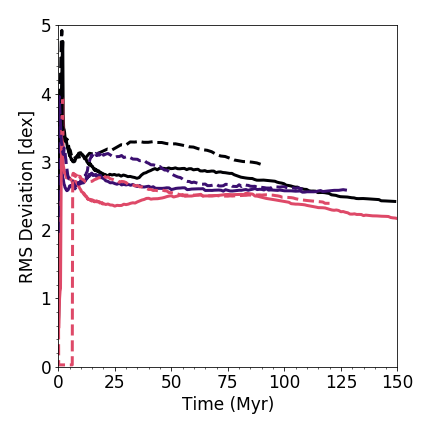}
  \caption{The same as Figure~\ref{fig:CGM_CNM} and Figure~\ref{fig:mean-median}, but comparing across both sets of runs with different SFRs. Line color corresponds to a fixed enrichment energy event, while the solid lines correspond to the higher SFR runs discussed throughout this work, and the dashed lines the lower SFR runs. See text for more details.}
  \label{fig:SFR_comparison_CGM_CNM}
\end{figure*}

Finally, we examine the companion run (\runtwonu) to see if the global SFR of the galaxy drives variation in these results. The correlation with SFR is of interest not because the formation of stars by themselves affects metal evolution, but rather the increase in feedback associated with a higher SFR (and thus a warmer / hotter ISM, greater turbulence in the ISM, and more significant outflows). We compare these two runs in Figure~\ref{fig:SFR_comparison_CGM_CNM}, showing runs from \runonenu~in as solid lines at higher concurrent SFR and runs at the lower SFR, \runtwonu, as dashed lines. $f_{\rm ej}$ increases much more rapidly across injection energies in \runonenu. However, this value seems to be similar across runs for E49 and E51 by $\sim$50 Myr, with an additional increase again in the \runonenu~simulations towards the end of the time. This initial spike, lull, and second increase corresponds to the period of active star formation, slight lull, and increase in SFR experienced during the \runonenu~simulations; the lull occurs between 40 and 80 Myr after the enrichment events. The SFR is consistently low (or zero) throughout \runtwonu. Although E49 and E51 exhibit similar behavior across global differences in SFR, the difference in E46 is significant throughout the examined time period. As these sources do not contain enough energy to eject their metals from the galaxy by themselves, they are only ejected from the galaxy by being swept up in the ISM during other feedback events. The evolution of these events are therefore much more dependent upon the global galaxy properties.

The fraction of metals contained within the CNM of the ISM does not seem to depend too strongly on the SFR, unlike $f_{\rm ej}$. It is generally true that after the initial $\sim$50 Myr of enrichment, a greater fraction of the metals in the ISM are contained in the CNM during the run with the lower SFR (\runtwonu), but this difference is not large and is only significant for the higher energy events. In addition, the mass fraction of the CNM is greater during run \runtwonu~than in run \runonenu, so it may simply be that the long-term evolution of the fraction of metals in the CNM is more dependent upon the phase structure of the ISM---which is regulated by stellar feedback---than the feedback directly.

Finally, examining the last two panels of Figure~\ref{fig:SFR_comparison_CGM_CNM}, it does appear that the mixing efficiency of metals for each source increases with increasing global SFR. In the initial evolution, all events in run \runtwonu~exhibit larger abundance spreads that take longer to begin mixing substantially than their counterparts in \runonenu. In general, \runtwonu~abundances remain larger for the first 100~Myr. The differences depend on the ejection energy, with the most significant difference found in comparing the E46 runs.

\section{Discussion and Conclusions}
\label{sec:discussion conclusion}

Galactic chemical evolution is far from single zone and metal mixing is clearly neither homogeneous nor instantaneous. As demonstrated in recent works \citep{Safarzadeh2017,Emerick2018b,KrumholzTing2018}, the properties of metal mixing in the ISM of various galaxies depends upon the characteristics of the individual sources. The energy of each source sets the thermal phase to which these metals couple to most effectively, and the volume over which they are initially injected before additional mixing by global galactic dynamics begins to dominate. In short, lower energy enrichment events, like those from AGB winds, mix much more slowly throughout the ISM than metals from higher energy events, like SNe. As found in \cite{Emerick2018b},
this difference is still present even when accounting for the fact that AGB events are both more common and evenly distributed in the galaxy than SNe.

We confirm the average trends found in previous works that sources with lower $E_{\rm ej}$ have larger abundance spreads in the ISM and lower ejection fractions, on average \citep{KrumholzTing2018,Emerick2018b}.%
\footnote{Interestingly, \cite{Safarzadeh2017} finds that the resulting abundance patterns in a UFD candidate at high redshift are generally insensitive to the energy of the injection event. However, they focus on enrichment events from a smaller range of injection energies: 10$^{50}$ - 10$^{51}$ erg.}
We additionally find that $E_{\rm ej}$ of the event and the global SFR during the time at which the event occurs produce the greatest variation in how many metals are retained by the galaxy, what fraction of those metals are contained within the CNM, and their homogeneity. Metals are in general ejected more effectively and mix more efficiently during periods of higher SFR than lower SFR, corresponding to periods of more effective galactic outflows and greater turbulence in the ISM. Although we have limited statistics to determine a conclusive trend, we generally find that neither the radial position of the event nor the local ISM density in which the event occurs has a significant effect on the average behavior of enrichment events, with the exception of the lower energy (E46) events.

 However, we find that the evolution of individual events can vary dramatically depending on the combination of each of these factors. This suggests that it would be challenging to make any conclusive statements about the enrichment behavior of single enrichment events. This is problematic for interpreting the chemical abundances in individual UFDs with abundance patterns that can be explained by a single, exotic enrichment event. Using these observations to constrain the total nucleosynthetic yields of these sources requires implicit assumptions about how quickly metals are available in star forming gas, how homogeneously they are distributed throughout the galaxy, and what fraction are ejected from the galaxy. It is likely that none of these quantities can be determined as simple parameters based upon the feedback properties of a single enrichment source or even the global properties of a galaxy. However, this study has shown that these parameters can be characterized in an averaged sense, even though the exact result is subject to substantial stochastic effects. Although interpreting the abundances of individual galaxies may be problematic, observations of stars across many UFDs with signatures of single, rare enrichment events might be leveraged to identify consistent trends. This, combined with one-zone models with paramaterization for these stochastic variations can be used to understand the possible outcomes for populations of these galaxies. Developing such a model will be a powerful tool for interpreting the irregular stellar abundance patterns observed in low mass dwarf galaxies in the nearby Universe.

Our results suggest that the spreads of stellar abundances at low metallicities may offer valuable insight into the origin of those metals. Lower energy events should generate larger abundances spreads than more energetic sources, as will particularly rare events. For example, $r$-process enrichment from exotic enrichment events like HNe should manifest itself as significantly more well-mixed than if the elements originated in lower energy NS-NS mergers. Unfortunately, the degeneracy between source energy and event frequency / distribution may make distinguishing different sources in observed stellar abundances challenging. However, the much more efficient expulsion of metals by high energy events might be able to break this degeneracy. Given these complications, turning these results into unique observational predictions requires further work. While the lowest mass, most metal poor dwarf galaxies are ideal environments to conduct this analysis, they typically only have a few to tens of measured stellar abundances. For better statistics, larger low-mass dwarfs, like Ursa Minor, Draco, Sextans, Sculptor, Carina and Fornax \citep[e.g][]{Suda2017,Duggan2018,Skuladottir2019}, and the low-metallicity stellar halo of the Milky Way \citep[e.g.][]{Hansen2018,Sakari2018} are likely the best regimes to try and use stellar abundances to infer the nucleosynthetic origin of elements and metal mixing properties in the ISM.

Our simulated galaxy lies in the regime where the turbulent driving is generated almost exclusively by supernova explosions (as opposed to gravitational driving from the inflow of gas, for example). Following the discussion in \cite{PanScannapiecoScalo2013}, mixing over the whole galaxy ($L_G$) must therefore occur slowly, as a random walk process of enrichment between individual, independent polluted regions in the galaxy on the transport timescale \begin{equation} \tau_{\rm trans} = L_{\rm G}^2 / (L_{\rm turb}v_{\rm rms}), \end{equation} 
where $L_{\rm turb}$ is the turbulent driving scale, comparable here to the typical size of a supernova remnant. Roughly, $L_{\rm G} \sim 1~{\rm kpc}$, $L_{\rm turb} \sim 100~{\rm pc}$, and $v_{\rm rms} \sim 10~{\rm km}~{\rm s}^{-1}$ for our galaxy, giving a mixing / transport timescale of $\sim$~1~Gyr. Our results, demonstrate that, for single enrichment sources, the mixing timescales are indeed quite long, with significant abundance variations remaining after 150~Myr of simulation time. For this reason, the frequency of enrichment events and their spatial distribution are key drivers of abundance homogenization in this regime, as demonstrated here. In addition, we find agreement with the analytic model in \cite{KrumholzTing2018} that the size of the initial enrichment region---which is directly correlated with injection energy---also determines the relative scatter in abundances throughout the ISM.

Current state-of-the-art semi-analytic models of galactic chemical evolution cannot readily capture the complex hydrodynamics effects governing the metal mixing process examined in this work. However, incorporating these effects could stand to dramatically improve the ability for these models to match not only the mean trends observed in stellar abundance patterns, but also their spreads. In turn, this could help better leverage these observations to understand both the nucleosynthetic origin of various elements, and also the properties of turbulent metal mixing operating within the ISM. We plan to incorporate the understanding gained from this work in such a model in future work.


\section*{Acknowledgments} We would like to thank Brian O'Shea, Benoit C{\^o}t{\'e}, Kathryn V.~Johnston, and Jason Tumlinson for valuable discussions and comments on a previous version of this work which appeared as a chapter in the first author's dissertation. In addition, we thank the anonymous referee whose comments have significantly improved this work. AE was supported by a Blue Waters Graduate Fellowship. GLB acknowledges support from NSF grants AST-1615955 and OAC-1835509 and NASA grant NNX15AB20G. M-MML was partly supported by NSF grant AST18-15461. We gratefully recognize computational resources provided by NSF XSEDE through grant number TGMCA99S024, the NASA High-End Computing Program through the NASA Advanced Supercomputing Division at Ames Research Center, Columbia University, and the Flatiron Institute. This work made significant use of many open source software packages. These are products of collaborative effort by many independent developers from numerous institutions around the world. Their commitment to open science has helped make this work possible. 

\software{\textsc{yt} \citep{yt}, \textsc{Enzo} \citep{Enzo2014}, \textsc{Grackle} \citep{GrackleMethod}, \textsc{Python} \citep{VanRossum1995python}, \textsc{IPython} \citep{perez2007ipython}, \textsc{NumPy} \citep{oliphant2006guide}, \textsc{SciPy} \citep{SciPy}, \textsc{Matplotlib} \citep{hunter2007matplotlib}, \textsc{HDF5} \citep{Fortner1998HDF,Koranne2011}, \textsc{h5py} \citep{h5py}, \textsc{Astropy} \citep{astropy:2013,astropy:2018}, \textsc{Cloudy} \citep{Cloudy2013}, and \textsc{deepdish}}




%
%

\bibliographystyle{yahapj}
\bibliography{refs}

\end{document}